\documentstyle[11pt,aas2pp4]{article}

\received{March 7 2002}
\accepted{May 1 2002}

\begin{document}

\title{The Effect of Radiative Cooling on the Scale-Dependence of
Global Stellar and Gas Contents of Groups and Clusters of Galaxies}

\author{Xiang-Ping Wu and Yan-Jie Xue}

\affil{National Astronomical Observatories, Chinese Academy
       of Sciences, Beijing 100012; and
       Institute of Astronomy and Astrophysics, Academia Sinica,
       Taipei 106, China}

\begin{abstract}
It is widely believed that the global baryon content 
and mass-to-light ratio of groups and clusters of galaxies
are a fair representative of the matter mix of the universe
and therefore, can be used to reliably determine the cosmic mass 
density parameter $\Omega_{\rm M}$. However, 
this fundamental assumption is challenged by growing evidence from 
optical and X-ray observations that the average gas mass 
fraction and mass-to-light ratio increase mildly with scale from 
poor groups to rich clusters. Although a number 
of time-consuming hydrodynamical simulations combined with 
semi-analytic approaches have been carried out, which permit 
a sophisticated treatment of some complicated processes in 
the formation and evolution of cosmic structures, the essential 
physics behind the phenomenon still remains a subject of intense 
debate. In this {\sl Letter}, using a simple analytic model, we show 
that radiative cooling of the hot intragroup/intracluster gas 
may allow one to reproduce the observed scale-dependence of 
the global stellar and gas mass fractions and mass-to-light ratio of 
groups and clusters, provided that about half of the cooled 
gas is converted into stars.
Together with the recent success in the recovery of 
the entropy excess and the steepening of the X-ray luminosity-temperature 
relations detected in groups and clusters,  radiative 
cooling provides a simple, unified scheme for the evolution of hot 
gas and the formation of stars in the largest virialized systems 
of the universe. 
\end{abstract}


Groups and clusters serve as a reservoir of baryons in the present
universe, which exist in the form of hot plasma with temperature 
close to the virial temperature ($10^6$-$10^8$ K) of the underlying
gravitational potential wells as a result 
of gravitationally-driven shocks and adiabatic compression
(Cen \& Ostriker 2000). 
The hot intragroup/intracluster gas would continuously lose energy 
due to thermal bremsstrahlung and line emissions. 
The decrease in X-ray temperature $T$ is completely governed  by 
the conservation of energy, $(3/2)nkT=\epsilon(n,T)t_{\rm cool}$, 
where $n$ is the total gas number density, and $\epsilon$ is the 
emissivity. This defines the so-called cooling time 
$t_{\rm cool}$ and radius $r_{\rm cool}$ within which gas can cool out of 
the hot phase. If the cooling time is set to equal the age of 
groups/clusters, or approximately the age of the universe, $t_0$,
we will be able to estimate the maximum cooling radii of the systems
by the present time, $r^{\rm m}_{\rm cool}$, and the corresponding critical 
gas density, $n(r^{\rm m}_{\rm cool})$. 
Since the pioneering work of White \& Frenk (1991), 
such a simple model for radiative cooling has been well incorporated in 
the study of formation and evolution of galaxies 
(e.g. Kauffmann, White \& Guiderdoni 1993; Kauffmann et al. 1999; 
Benson et al. 2000; Wu, Fabian \& Nulsen 2000; 2001; 
Somerville et al. 2001; Balogh et al. 2001;  Yoshida et al. 2002; etc.).

We assume that in the absence of cooling the gas has the same 
distribution as the dark matter in groups and clusters.
We adopt the universal density profile 
$\rho_{\rm NFW}(r)\propto 1/[cr(1+cr)^2]$ suggested by 
numerical simulations for the dark halos of groups/clusters
(Navarro, Frenk \& White 1997)
and specify the concentration parameter $c$ by
$c=10(M/2.1\times10^{13}M_{\odot})^{-0.14}$ for a given halo 
of mass $M$ (Bullock et al. 2001). 
The number density of the hot gas thus follows 
$n(r)=(f_{\rm b}/\mu m_{\rm p})\rho_{\rm NFW}(r)$, where $f_{\rm b}$ is 
the universal baryon fraction, and $\mu=0.59$ is the mean molecular weight. 
We assign an X-ray temperature to each halo in terms of cosmic virial theorem,
$M\propto T^{3/2}$ (Bryan \& Norman 1998).

The gas within the maximum cooling radius $r^{\rm m}_{\rm cool}$ 
is assumed to convert into stellar objects (Pearce et al. 1999; Bryan 2000). 
Note that this latter component should also include 
other possible cooled materials  
(e.g. neutral and molecular gas) which may form out of cooling process.  
The cooled gas mass within $r^{\rm m}_{\rm cool}$
can be obtained by integrating the gas 
profile $n(r)$ over volume out to $r^{\rm m}_{\rm cool}$.
Actually, such a simple exercise gives rather a robust estimate of the total 
stellar mass of a system, $M_{\rm star}$, as has been shown recently by
Yoshida et al. (2002).
The gas outside $r^{\rm m}_{\rm cool}$ may also give some contribution 
to $M_{\rm star}$. This happens because the central region 
can be refilled with the hot gas distributed originally 
at large radii due to the lack of pressure support once
the gas within $r^{\rm m}_{\rm cool}$ cooled out of the hot phase.
Here we use a less vigorous method to estimate the amount of 
this secondary cooled component due to successive cooling.
We first work out the newly established gas distribution by combining the 
conservation of entropy and the equation of hydrostatic 
equilibrium (Bryan 2000; Voit \& Bryan 2001; Wu \& Xue 2002).
We then play the same game as the above by setting 
the age of the universe to equal the cooling time, which defines the 
secondary cooing radius and thus yields the total mass of corresponding
cooled material. 
Numerical computation shows that the contribution of 
this inward flow to the total stellar mass is less than 
$20\%$ in groups and clusters. 
Finally, we obtain the stellar and gas mass fractions from 
$f_{\rm star}=M_{\rm star}/M$ and $f_{\rm gas}=f_{\rm b}-f_{\rm star}$,
respectively. Because the hot gas cools relatively faster in groups than
in clusters due to the difference in their density contrasts and/or
temperatures, 
poor groups experienced higher efficiency of star formation
than rich clusters did. Therefore, a(n) decrease (increase) of stellar (gas) 
mass fraction from groups to clusters is expected to occur 
naturally (Pearce et al. 1999).
A long standing question is:  
Can the expected scale-dependence of $f_{\rm star}$
and $f_{\rm gas}$ as a result of radiative cooling 
be quantitatively reconciled with observations ?

We take a flat, cosmological constant dominated cosmological model 
to proceed our numerical calculation: $\Omega_{\rm M}=0.3$ and 
$\Omega_{\Lambda}=0.7$, in which the baryon density is 
$\Omega_{\rm b}=0.047$  
for a Hubble constant of $H_0=65$ km s$^{-1}$ Mpc$^{-1}$. The present
age of the universe is correspondingly $t_0=14.5$ Gyr. 
We employ an optically thin and isothermal plasma emission model
with a primordial mixture of $77\%$ hydrogen and $23\%$ helium 
and a metallicity of $30\%$ solar (Raymond \& Smith 1977).
In Figure 1 we compare our predicted stellar and gas mass fractions 
as a function of X-ray temperature $T$ with three optical and X-ray 
observations of 33 groups and clusters (Mulchaey et al. 1996;
Cirimele, Nesci \&  Trevese 1997;  Hwang et al. 1999). 
These samples were used as an input 
in the galaxy formation-regulated gas evolution model aimed at
the explanation of the observed entropy floor in the central cores
and X-ray luminosity distribution of groups and clusters
(Bryan 2000; Wu \& Xue 2002).
Because the theoretical predictions of $f_{\rm star}$ and $f_{\rm gas}$ 
depend critically on the truncated radii of groups and clusters, 
we demonstrate the results for two choices of 
overdensity parameter, $\Delta=200$ and $2500$, which
cover roughly the observational range according to the 
calibration of Evrard, Metzler \& Navarro (1996).
While there is essentially a good agreement 
between the theoretical predictions and the observations, 
the observed data within different radii 
which increase roughly from groups (larger $\Delta$) to rich clusters 
(smaller $\Delta$) may also introduce a systematic bias giving rise 
to the similar scale-dependence. In order to eliminate this concern,
we turn to the sample of Roussel, Sadat \& Blanchard (2000) who derived
the stellar and gas mass fractions of 33 groups and clusters out to
the virial radii. Of course, their result  should also be used with
caution because the evaluation of the total stellar and gas mass fractions
requires a substantial extrapolation of the observational data 
to virial radii especially for groups and poor clusters.
Recall that the $\beta$
parameters in the X-ray surface brightness profiles of groups and poor
clusters revealed by current observations are often smaller than $\sim0.6$,
yielding a monotonically increasing gas mass fraction with radius
(Wu \& Xue 2000). This may lead to a significant overestimate
of the total gas mass fractions of groups and poor clusters.
The observed data of Roussel et al. (2000) are illustrated in Figure 2,
together with our predictions in terms of radiative cooling.  
Although  Roussel et al. (2000) didn't found any strong evidence 
for a positive correlation between gas mass fraction and temperature, 
their data, nevertheless, are still  consistent with the mild increase 
of $f_{\rm gas}$ with
$T$ expected from the cooling model. However, the stellar mass fractions
in most of the groups and clusters are smaller than the theoretical 
prediction by a factor of $\sim2$.  This discrepancy  
indicates that a considerably large fraction of the cooling gas 
in groups and clusters may not be converted 
into stellar objects. Instead, the cooling gas may end up in other forms
such as cold clouds.  Alternatively, the discrepancy seems 
to be more significant on group scales than on cluster scales. 
Namely, the theoretically predicted scale-dependence of $f_{\rm star}$ 
is slightly stronger than the observation. Yet,  
the observations are at present uncertain, so that such 
a claim may still be premature.

It follows immediately that 
the dependence of the stellar mass fraction of groups and clusters
on X-ray temperature would result in a variation of mass-to-light 
ratio from groups to clusters. This happens because the mass-to-light
ratio can be formally written as 
$M/L=(M/M_{\rm star})(M_{\rm star}/L)=\Upsilon/f_{\rm star}$, where 
$\Upsilon\equiv M_{\rm star}/L$ measures the efficiency with which
groups and clusters transform cooled material into light.
Once this free parameter is specified,  
we will be able to predict how $M/L$ varies with $M$ or $T$.

Bahcall \& Comerford (2002)  have compiled a sample of 21 systems 
whose X-ray temperatures and mass-to-light ratios are reliably determined.
Over the temperature range from groups ($kT=0.73$ keV) to rich clusters 
($kT=12.3$ keV), they have found that $M/L_V$, which are corrected
to redshift $z=0$, show an increasing tendency toward high temperature
clusters. In Figure 3 we illustrate their observed $M/L_V$ versus $kT$,  
together with our theoretical predictions for 
$\Upsilon=7.5\Upsilon_{\odot}$. 
It turns out that the two results are reasonably 
consistent with each other. The slightly large value of 
$\Upsilon=7.5\Upsilon_{\odot}$ as compared with that 
($\approx5\Upsilon_{\odot}$) of early-type galaxies 
(Fukugita, Hogan \& Peebles 1997) may arise from the fact that 
our total stellar mass $M_{\rm star}$ from cooling model has also 
included other cooled components (e.g. cold clouds).  
However, there exist two biases which may invalidate our comparison: 
The observed data were taken within different radii of 
groups and clusters, which may lead to an increasing $M/L$ with $T$ 
if the X-ray measurements had a bias toward the selection of central
X-ray luminous regions for low-mass systems (groups)
but more extended regions for rich clusters. 
Another bias factor arises from the fact that $\Upsilon$ itself 
may have a positive correlation with $M$ or $T$ (Balogh et al. 2001),
in the sense that groups contain mainly spirals 
($M/L_B\approx1.5\Upsilon_{\odot}$) while rich clusters are dominated 
by ellipticals ($M/L_B\approx6.5\Upsilon_{\odot}$)
(Fukugita et al. 1997).

Next, we turn to the hitherto largest sample of $M/L_B$ for groups and
clusters compiled recently by Girardi  et al. (2002).  
We restrict ourselves to
the massive systems with $M\geq10^{13}M_{\odot}$ to ensure
that the dominant X-ray emission is produced by `primordial' gas rather
than stellar winds, supernova remnants or star binaries. 
This reduces to a subsample of 213 systems (see Figure 4). 
Meanwhile, both $M$ and $M/L_B$ of all the systems 
have been computed to their virial radii, which eliminates at least 
one of the concerns addressed above. Yet, the data set has its own 
problem:  The two variables, $M/L_B$ and $M$, are not independent 
of each other. This arises from the fact that in their sample 
the uncertainty in $M$ is larger than that in $L_B$, and the error 
ellipses introduce an intrinsic, positive 
correlation between the two quantities.
In Figure 4 we demonstrate our predicted $M/L_B$ versus $M$ 
assuming $\Upsilon=8.5\Upsilon_{\odot}$. 
The adoption of $\Upsilon=8.5\Upsilon_{\odot}$ in the B band required 
to match the observed $M/L$ distribution is
compatible with $\Upsilon=7.5\Upsilon_{\odot}$ in the V band used in 
Figure 3, if we make use of the synthetic model by Charlot et al. (1996)
to calculate the $B-V$ color for a galaxy age of $\sim12$ Gyr.
The larger $\Upsilon$ parameter relative to the mean value
of  $\Upsilon\approx4.5\Upsilon_{\odot}$ for a typical cluster 
(Balogh et al. 2001) implies that about half of 
the cooled gas deposited from cooling flows may condense 
into other form of cold materials in the central regions of groups 
and clusters. This conclusion is consistent with the existence
of the discrepancy between the predicted and observed cooled 
gas components shown in Figure 2.

We can also evaluate the mass deposition rate $\dot{M}_{\rm cool}$ within 
cooling radius and its cosmic evolution: 
$\dot{M}_{\rm cool}=4\pi(f_b/\mu m_{\rm p})
\rho_{\rm NFW}(r_{\rm cool})r_{\rm cool}^2\dot{r}_{\rm cool}$,
where $\dot{r}_{\rm cool}$ is determined by combining
the conservation of energy with the self-similar model (NFW) for gas.
It appears that if about half of the cooling gas is converted into stars, 
our estimated $\dot{M}_{\rm star}$ at present epoch varies from 
$10$ $M_{\odot}$/yr for groups of $M=10^{13}$ $M_{\odot}$ 
to $500$ $M_{\odot}$/yr 
for clusters of $M=10^{15}$ $M_{\odot}$. This indicates  
substantial and on-going star formation in today's rich clusters. 
The detection of anomalously blue colors in the central regions of 
clusters (Allen 1995) does give support to a 
high rate of recent star formation, though a quantitative comparison
with observations is hampered by large uncertainty in the current 
derived star formation rate in clusters. Recent Chandra observations
of several rich clusters have revealed that the integrated mass deposition 
rates are about $100$ -- $500$ $M_{\odot}$/yr 
within the cooling radii ($\sim100$ kpc) and drop to  
$\sim10$ $M_{\odot}$/yr in the very central regions ($\sim10$ kpc)
(Allen et al. 2001; Allen, Ettori \& Fabian 2001; 
Schmidt, Allen \& Fabian 2001; Ettori et al. 2002; 
etc.).  Alternatively,  our predicted stellar mass fraction 
and star formation rate show no significant change 
between nearby and distant groups/clusters out to redshift $z\approx1$
if they have the same mass, implying an
early formation of stars in groups and clusters and little cosmic
evolution of their stellar mass fractions and mass-to-light ratios
within $z\approx1$. 
Note that the concentration parameter $c$ of a distant dark halo at
redshift $z$ decreases by approximately a factor of ($1+z$) 
as compared with that of 
a nearby halo with identical mass (Navarro et al. 1997; Bullock et al. 2001).
Namely, while the young 
age of high-redshift clusters is unfavorable for the accumulation of 
cooled material, the dense matter environment of the clusters 
gives rise to a relatively short cooling time. The combined effect 
accounts for the lack of significant cosmic evolution 
of $f_{\rm star}$, $f_{\rm gas}$ and $M/L$ since $z\approx1$.

We have so far concentrated on the global properties of groups and 
clusters as a consequence of radiative cooing. In fact, 
the regulated gas distribution in groups and clusters produced by 
cooling or star formation resembles the conventional 
$\beta$ model in shape (Wu \& Xue 2002).
In particular, the scale-dependence of
radiative cooling or star formation process explains naturally
the entropy excess and the steepening of the X-ray 
luminosity-temperature relations in groups and clusters
(Bryan 2000; Pearce et al. 2000; Muanwong et al. 2000; 
Voit \& Bryan 2001; Wu \& Xue 2002). 
In a word, radiative cooling of the hot intragroup/intracluster gas,
which is based on the well-motivated physical process, may allow
us to resolve all the puzzles seen in current X-ray observations
of groups and clusters, and energy feedback from star formation
comes into effect only in the less massive systems of 
$M<10^{13}$ $M_{\odot}$.

Apparently, the present simple model has its own problems that 
need to be resolved in the future. First, the cosmic evolution
of groups and clusters has not been included, which is relevant to
the well-known overcooling crisis if massive dark halos like groups
and clusters form by gravitational aggregation of individual low-mass 
galaxies.
Our analytic cooling model is only applicable to the case 
when most of the hot gas had already assembled in groups and clusters.
In other words, we have attempted to address the question of how much
the hot gas heated by gravitational shocks and adiabatic compression
in groups and clusters has cooled out of the hot phase
by the present epoch. Second, our model without the inclusion of
energy feedback from star formation becomes invalid for low-mass
systems of $M<10^{13}$ $M_{\odot}$, because the hot gas 
can be expelled by supernova explosion from the shallow gravitational
potentials of these systems. In effect, our simple model predicts that 
radiative cooling alone may have consumed most of the hot gas in galaxies 
with masses below $10^{12}$ $M_{\odot}$. This is at odds with the
recent estimate of the global fraction of baryons which have cooled
by now (Balogh et al. 2001). In addition, the cooling model seems
to yield somewhat a stronger scale-dependence of cooled gas component 
than both observations (Figure 2) and hydrodynamical simulations 
(Figure 4), regardless of the discrepancy of a factor of $\sim2$ 
in amplitude. This could also be attributed to the influence of star 
formation. Indeed, energy feedback can heat 
the intragroup/intracluster medium,
which is equivalent to suppressing the cooling efficiency. The fact
that the effect is more significant in groups than in clusters may
suppress moderately the scale-dependence of the cooled gas content 
predicted by radiative cooling.  Third, the main reason behind
radiative cooling for the  scale-dependence of stellar and gas 
contents of groups and clusters remains unclear. Recall that 
the present investigation is based solely on the conservation of energy, 
$(3/2)nkT=\epsilon(n,T)t_{\rm cool}$. Either higher gas density ($n$), 
or lower temperature ($T$), or less thermal energy [$(3/2)nkT$] in
low-mass systems (e.g. poor clusters and groups) should be able to
account for the scale-dependence of stellar and gas contents from
groups to clusters. Future work will thus be needed to clarify the issue.

\acknowledgments
We gratefully acknowledge the constructive suggestions by an anonymous
referee. This work was supported by 
the National Science Foundation of China, 
the Ministry of Science and Technology of China, under Grant
No. NKBRSF G19990754, and 
the National Science Council of Taiwan, under Grant NSC91-2816-M001-0003-6.


\clearpage

\figcaption{Gas mass fraction $f_{\rm gas}$ (solid lines) and  
stellar mass fraction $f_{\rm star}$ 
(dashed lines) as a function of temperature $T$ predicted by 
radiative cooling are compared with  
the observed $f_{\rm gas}$ (filled circles) and $f_{\rm star}$ 
(open squares) of 33 groups and clusters
(Mulchaey et al. 1996; Cirimele et al. 1997;  Hwang et al. 1999).
Since the observational results were obtained within different radii 
corresponding roughly to overdensity parameter ranging from 200 to 2500
in different groups and clusters, the theoretical predictions are plotted
for two choices of overdensity parameter, 
$\Delta=200$ and $2500$, respectively. 
\label{fig1}}

\figcaption{Total gas mass fraction (solid line) and stellar mass 
fraction (dashed line) within virial radius predicted by radiative 
cooling are compared  with the observed data of 33 groups and clusters 
by Roussel et al. (2000): filled circles - $f_{\rm gas}$, and 
open squares - $f_{\rm star}$. 
Dotted line represents the result when  
the cooled gas mass fraction is reduced by a factor of 2.
\label{fig2}}

\figcaption{The dependence of mass-to-light ratio on X-ray temperature 
is shown for overdensity parameter $\Delta=200$ (upper line) and  
$\Delta=2500$ (lower line) assuming a cooled gas mass-to-light
ratio of $\Upsilon=7.5\Upsilon_{\odot}$ in the V band.
The data points of 21 systems from Bahcall \& Comerford (2002)
measured within radii of typically 0.8 to 2.3 Mpc fall essentially 
within the theoretical predictions. 
\label{fig3}}

\figcaption{The dependence of mass-to-light ratio $M/L$ on virial mass $M$. 
The observed data points of 213 groups and clusters (points) 
are taken from Girardi et al. (2002). Dashed line is our predicted $M/L$ 
distribution for an overdensity of $\Delta=200$ and 
$\Upsilon=8.5\Upsilon_{\odot}$ in the B band. 
Solid lines represent the results obtained by numerical simulations
(from top to bottom: Somerville et al. 2001;  Benson et al. 2000; 
Kauffmann et al. 1999), in which
the mean mass-to-light ratio of the universe has been assumed
to be $878$ $M_{\odot}/L_{\odot}$ for the data of 
Kauffmann et al. (1999). 
\label{fig4}}

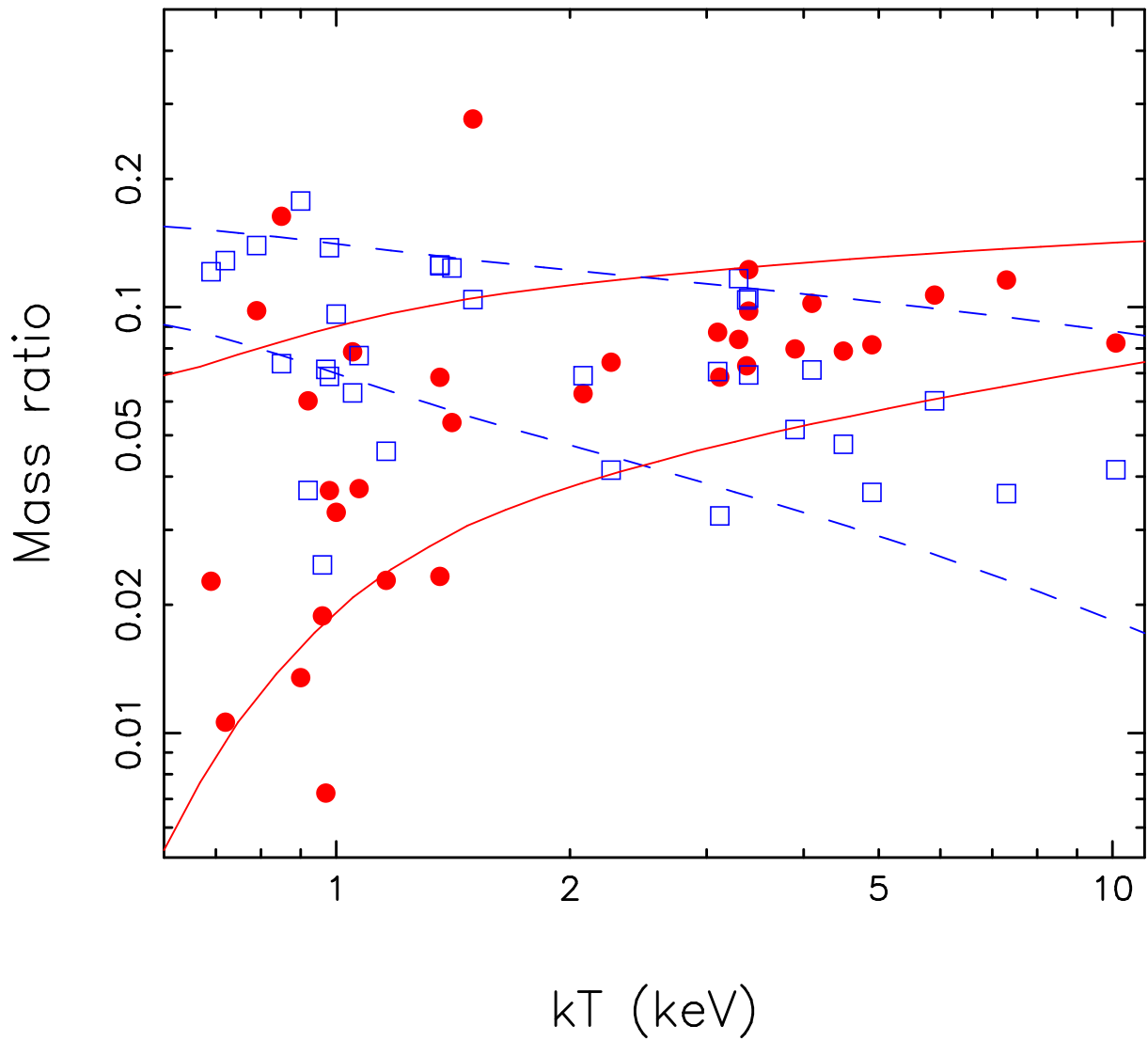
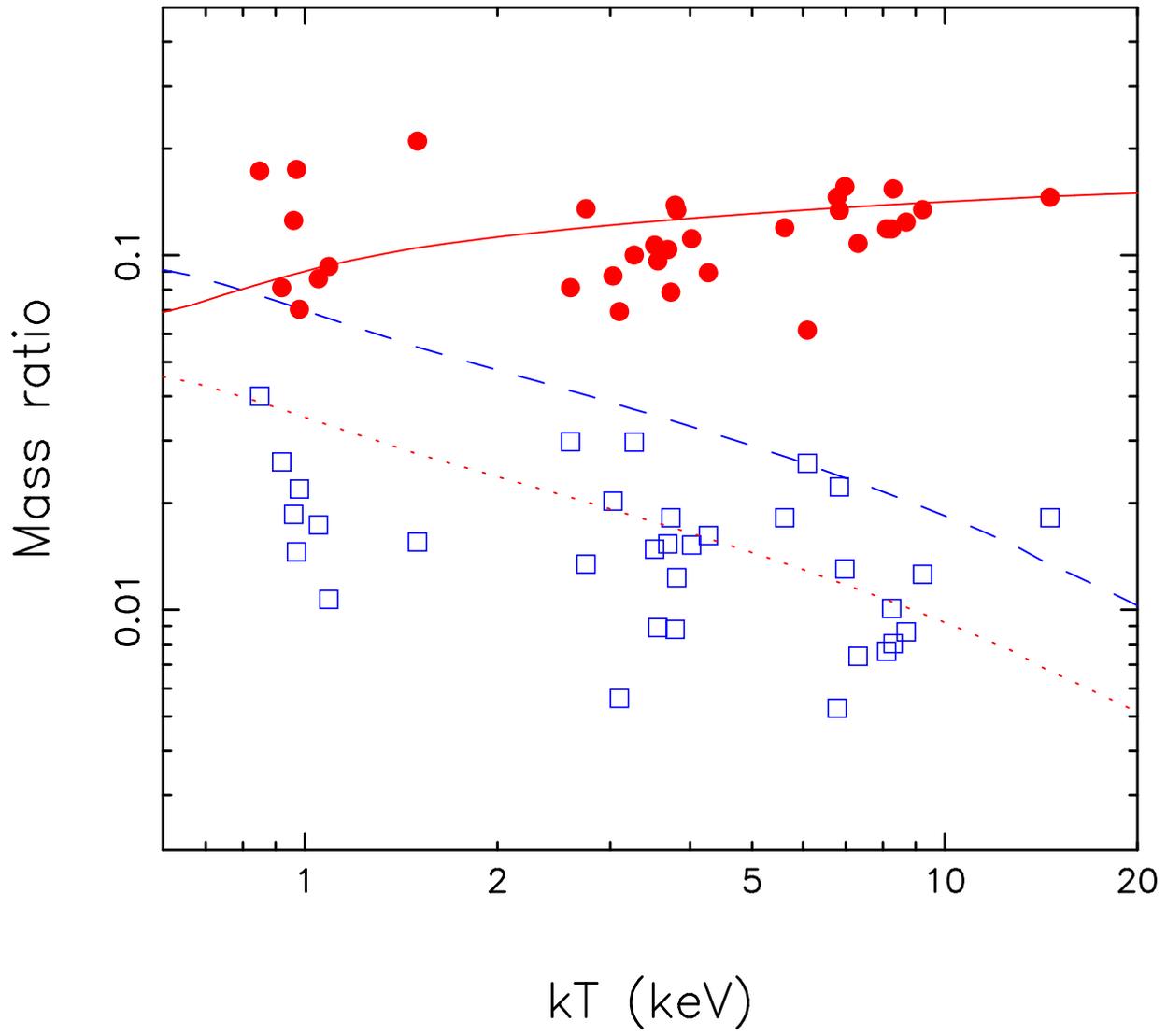
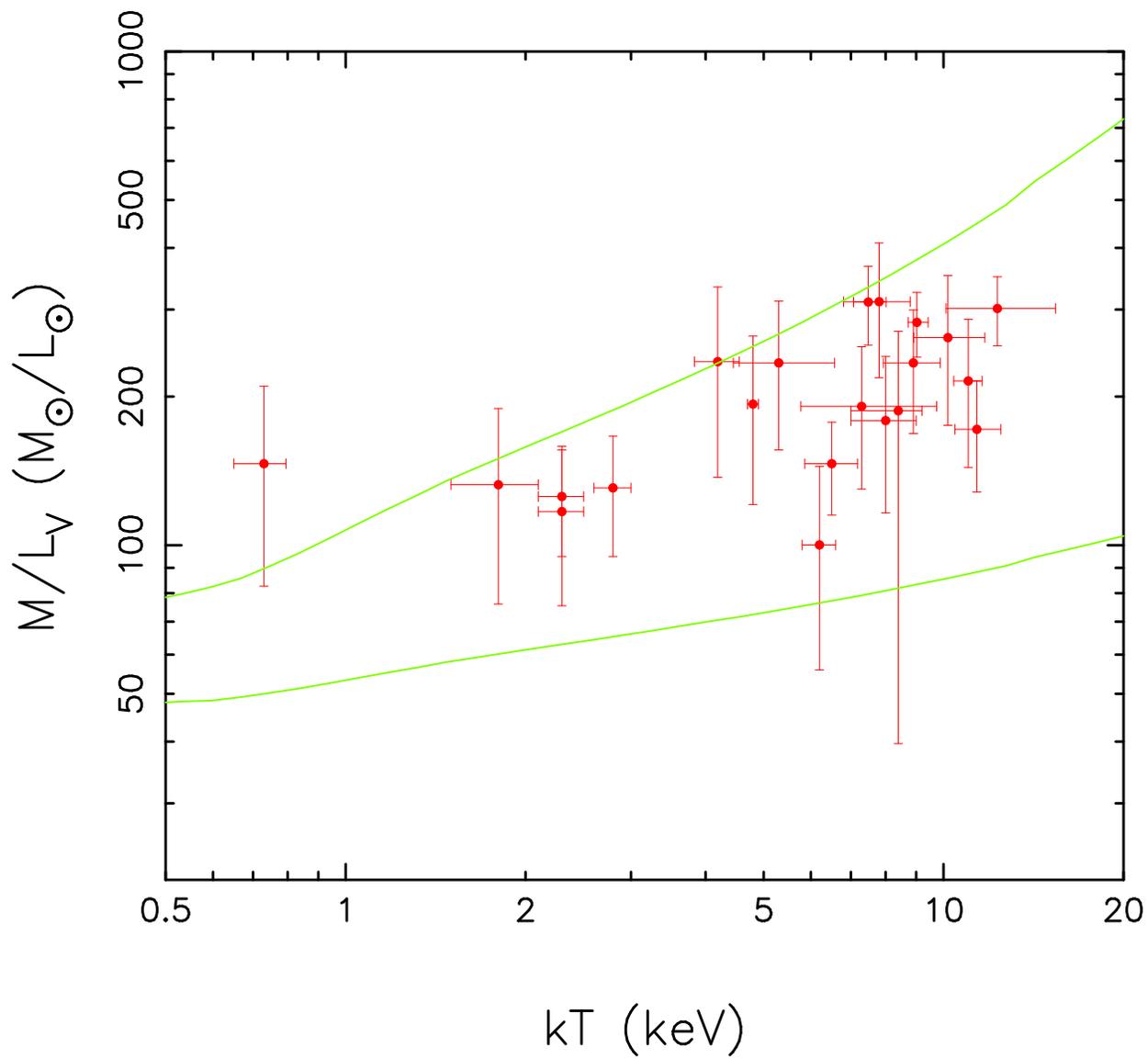
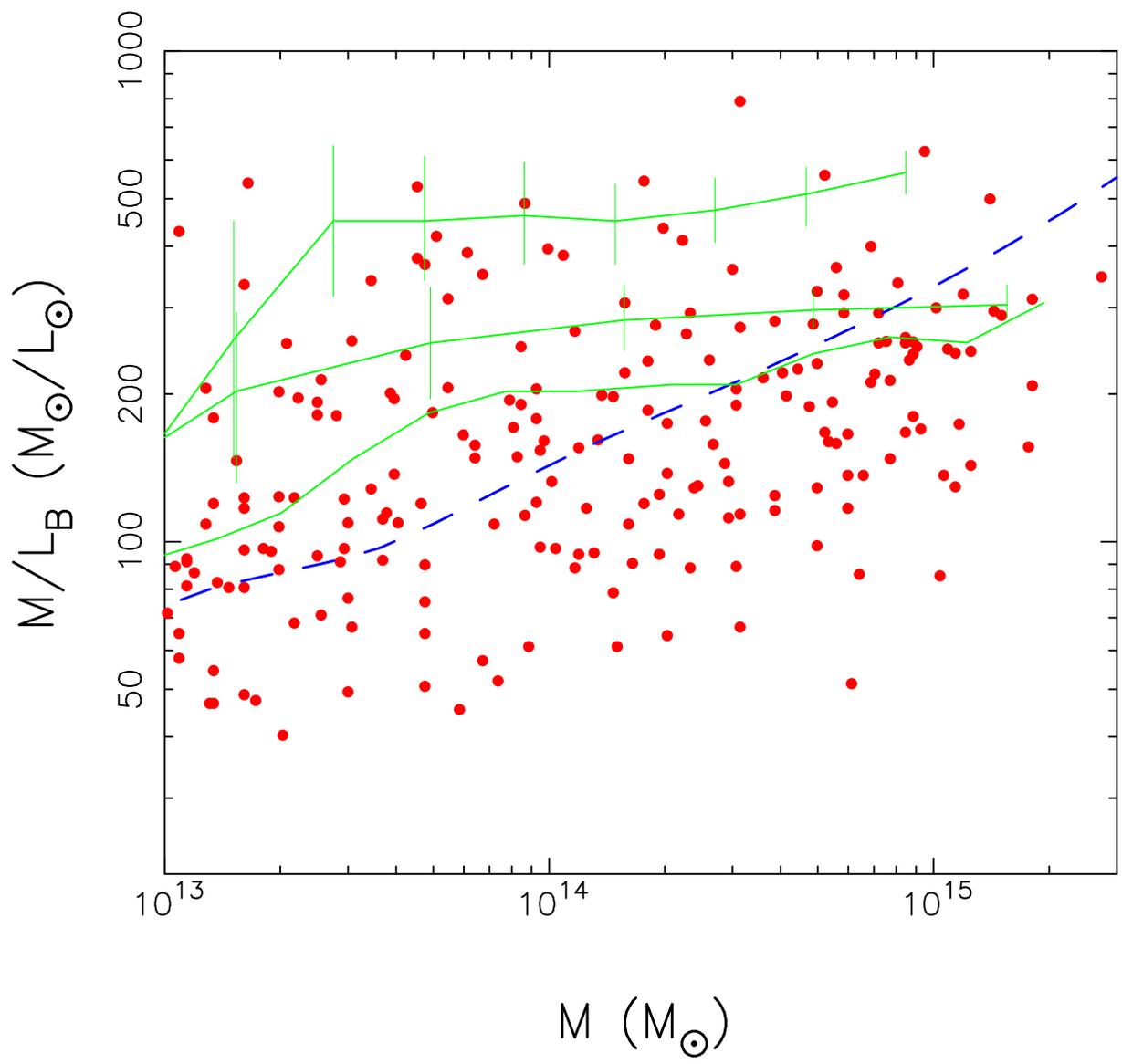
\end{document}